\documentclass[10pt,conference,final,twocolumn]{IEEEtran}

\usepackage{color}
\usepackage{bm}
\usepackage{bbm}
\usepackage{dsfont}
\usepackage{lipsum}

\usepackage[caption=1]{caption}
\usepackage{float}
\usepackage{stfloats}
\usepackage{eqparbox}
\usepackage{indentfirst}

\usepackage{fancyhdr}
\usepackage[style=ieee,maxnames=4,minnames=3,maxbibnames=3]{biblatex}

\usepackage{makeidx}

\usepackage{amsfonts}
\usepackage[cmex10]{amsmath}
\usepackage{amsthm}
\usepackage{amssymb}
\usepackage{mdwmath}
\usepackage{mathtools}
\usepackage{cases}

\theoremstyle{definition}

\theoremstyle{remark}

\theoremstyle{remark}

\theoremstyle{remark}

\theoremstyle{remark}

\usepackage{tabularx}
\usepackage{array}
\usepackage{mdwtab}

\usepackage{algorithm}
\usepackage{algpseudocode}

\usepackage[pdftex]{graphicx}
\usepackage{epstopdf}
\usepackage{subcaption}
\graphicspath{ {./img/} }

\usepackage{url}

\begin{document}

\title{Physics-Inspired Target Shape Detection and Reconstruction in mmWave Communication Systems \vspace{-0.1cm}}
\author{Ziqing~Xing\textsuperscript{$\dagger$}, Zhaoyang~Zhang\textsuperscript{$\dagger$}, Xin~Tong\textsuperscript{$\dagger$}, Zhaohui~Yang\textsuperscript{$\dagger$}\textsuperscript{$\ddagger$},   
Chongwen~Huang\textsuperscript{$\dagger$}\\ 
    \IEEEauthorblockA{\textsuperscript{$\dagger$}College of Information Science and Electronic Engineering, Zhejiang University, Hangzhou, China\\
        \textsuperscript{$\dagger$}Zhejiang Provincial Key Laboratory of Info. Proc., Commun. \& Netw. (IPCAN), Hangzhou, China\\
        \textsuperscript{$\ddagger$}Zhejiang Lab, Hangzhou, China\\
        E-mails: \{ziqing\_xing, ning\_ming, tongx, yang\_zhaohui, chongwenhuang\}@zju.edu.cn \vspace{-0.4cm}
    }
}

\date{}
\maketitle

\begin{abstract}
The integration of sensing and communication (ISAC) is an essential function of future wireless systems. 
Due to its large available bandwidth, millimeter-wave (mmWave) ISAC systems are able to achieve high sensing accuracy. 
In this paper, we consider the multiple base-station (BS) collaborative sensing problem in a multi-input multi-output (MIMO) orthogonal frequency division multiplexing (OFDM) mmWave communication system. Our aim is to sense a remote target shape with the collected signals which consist of both the reflection and scattering signals. We first characterize the mmWave's scattering and reflection effects based on the Lambertian scattering model. Then we apply the periodogram technique to obtain rough scattering point detection, and further incorporate the subspace method to achieve more precise scattering and reflection point detection. Based on these, a reconstruction algorithm based on Hough Transform and principal component analysis (PCA) is designed for a single convex polygon target scenario. To improve the accuracy and completeness of the reconstruction results, we propose a method to further fuse the scattering and reflection points. Extensive simulation results validate the effectiveness of the proposed algorithms. 
\end{abstract}

\begin{IEEEkeywords}
Integration of sensing and communication, millimeter wave, channel modeling, target shape reconstruction \vspace{-0.1cm}
\end{IEEEkeywords}

\section{Introduction}
\vspace{-0.1cm}
Future sixth-generation (6G) wireless communication system will drive various emerging applications, 
such as smart cities, Industrial Internet of Things, autonomous driving, etc \cite{cui2021integrating, xu2023edge}. 
These potential applications will undoubtedly require high-quality wireless connections and high-precision sensing capabilities at the same time. 
Therefore, the integration of sensing and communication (ISAC) \cite{ISAC_liu2022} will become an essential enabling technology, 
which aims to provide both traditional communication services and environmental sensing services simultaneously. 
It can not only save resource expenses but also enable mutual benefits between the communication and sensing functionalities \cite{JCAR_zhang2021, xiong2023fundamental}.

Compared with the sub-6GHz frequency band, the millimeter wave (mmWave) has a shorter wavelength and larger bandwidth, 
which make it more promising for the ISAC system. 
In order to fully utilize the potential function in sensing, establishing a mmWave channel model is a crucial step.
Currently, most traditional channel models are based on the statistical channel characteristics, 
such as power-delay and power-angle profiles \cite{mmWave_hemadeh2017}. 
However, statistical channel models cannot provide accurate channel parameters for specific scenarios, 
which is fatal for environment sensing tasks. 
Recently, ray-tracing \cite{RT_yun2015} method has been applied in communication channel simulation to establish more accurate deterministic channel models. 
Nonetheless, the work in \cite{RT_yun2015} mainly considered reflection paths, while ignoring the scattering effects of mmWaves, which is not conducive to extracting detailed environmental information. 
How to construct a more comprehensive ray-tracing model to serve radio frequency sensing is still a research topic worthy of exploration.

Computational imaging is an emerging environment sensing method. 
The authors in \cite{BSBL_yao2021} divided the sensing scenario into voxels and proposed an on-grid compressed sensing algorithms for environment reconstruction based on the block sparsity of the environment. 
Furthermore, a multi-view sparse vector reconstruction algorithm was proposed to solve the occlusion problem \cite{MVSVR_tong2022}. 
However, for large-scale environment, such dense grid partition in \cite{BSBL_yao2021} and \cite{MVSVR_tong2022} can cause an unacceptable computation overhead. 
The periodogram and the subspace are widely-used target parameter estimation methods in radar signal processing, 
which exploit the system's degrees of freedom in the time-frequency-space domain to efficiently estimate continuous target parameters, such as azimuth, range, velocity, etc \cite{MIMO_OFDM_location_sun2018}. 
These classic algorithms do not require grid partitioning of the environment and have a lower computation cost. 
However, the conventional radar signal processing methods, such as periodogram and subspace, cannot be directly used for accurate environment sensing since these methods lack the utilization of the overall physical features of the environment. 

In this paper, we first model the physical process of mmWave propagation. 
Then, based on the obtained signal at the receiver, we design a scattering and reflection point detection algorithm to acquire the target points in the ray tracing model by combining periodogram and subspace methods.
Based on the calculated scattering and reflection points, we further propose a sensing target shape reconstruction algorithm. 
The main contributions of this paper are summarized as follows:
\begin{itemize}
    \item We analyze the mmWave propagation progress based on ray-tracing, and design a scattering and reflection point sensing algorithm by combining the periodogram and subspace methods. 
    \item For a single convex polygon target, we design a Hough transform and principal component analysis based target shape reconstruction (HT-PCA-TSR) algorithm, and further refine the accuracy and completeness of the reconstruction results by fusing scattering and reflection points.
    \item Substantial simulation results validate the effectiveness of the proposed target sensing algorithm.
\end{itemize}

\section{System Model}\label{sec::system_model}
\subsection{Multi-BSs Sensing Scenario}\label{subsec::scenerio}
\begin{figure} [!htbp] 
\vspace{-0.3 cm}
\centering
\includegraphics[width=\linewidth]{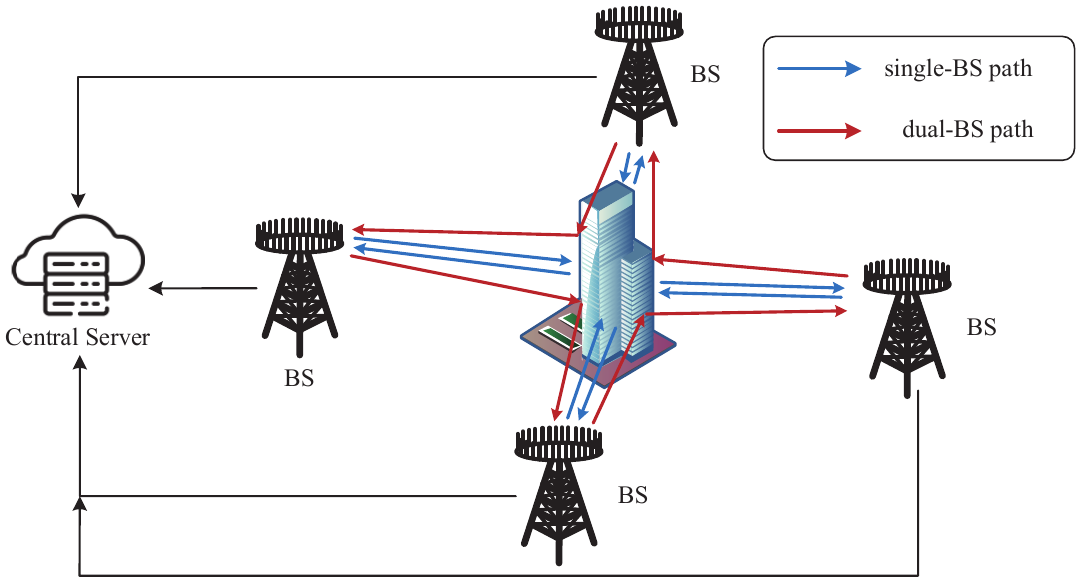}
\vspace{-0.4 cm}
\caption{Considered multi-BS environment sensing scenario.}
\label{fig::scenerio}
\vspace{-0.3 cm}
\end{figure}

We consider a multi-base stations (BSs) environment sensing scenario in a multiple-input multiple-output (MIMO) orthogonal frequency division multiplexing (OFDM) system, as shown in Fig. \ref{fig::scenerio}. 
In the considered scenario, single-BS means that the BS receives the reflection signals of self-sending signals, while dual-BS stands for that the BS receives the signal from neighboring BSs. 
Each BS periodically sends pilots and receives signals from itself and neighboring BSs 
to obtain the channel state information (CSI) of single-BS paths and dual-BS paths. 
We use the time division method to coordinate the active sensing phase of different BSs to avoid mutual interference.
Each BS senses the local view of the environment and uploads the results to the central server for multi-view process. 

This paper focuses on the shape sensing of objects on a two-dimensional plane, 
and specifically analyzes the case of a single convex polygon target.
We assume that every BS uses a uniform linear array (ULA), and each array element is an omnidirectional antenna.

\subsection{Signal Propagation and Channel Model}\label{subsec::channel_model}
Ray tracing is a deterministic channel modeling tool that can provide parameters in specific scenarios, 
including path loss, delay, angle of arrival (AoA), and angle of departure (AoD). 
The shooting and bouncing ray (SBR) method is a way to implement ray-tracing, 
which discretizes the radiation field of the transmitter in the angular domain to approximate signal propagation with ray clusters. 

According to \cite{mmWave_hemadeh2017}, we consider the reflection and scattering of the mmWave on the object surfaces. 
For a single convex polygon target, we only need to consider the first-order reflection or scattering path, 
since the line-of-sight (LOS) path can be eliminated based on the BS location.

\begin{figure}[!htbp]
    \vspace{-0.3 cm}
    \centering
    \begin{subfigure}[b]{0.49\linewidth}
        \centering
        \includegraphics[width=\linewidth]{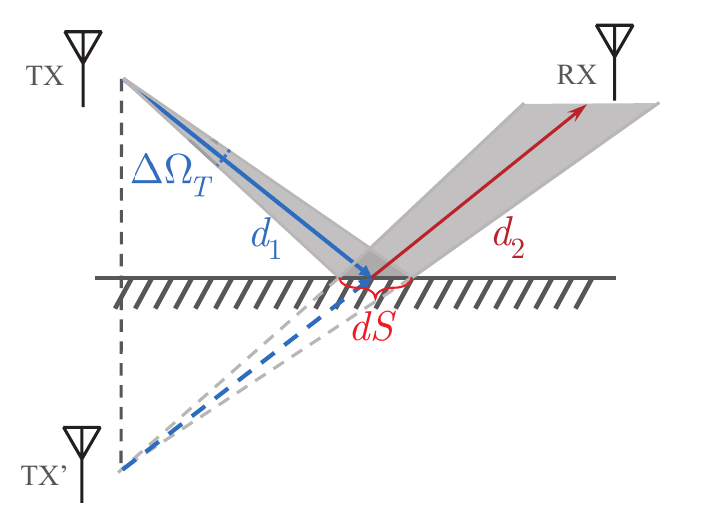}
        \caption{The reflection model.}
        \label{fig::reflection}
    \end{subfigure}
    \hfill
    \begin{subfigure}[b]{0.49\linewidth}
        \centering
        \includegraphics[width=\linewidth]{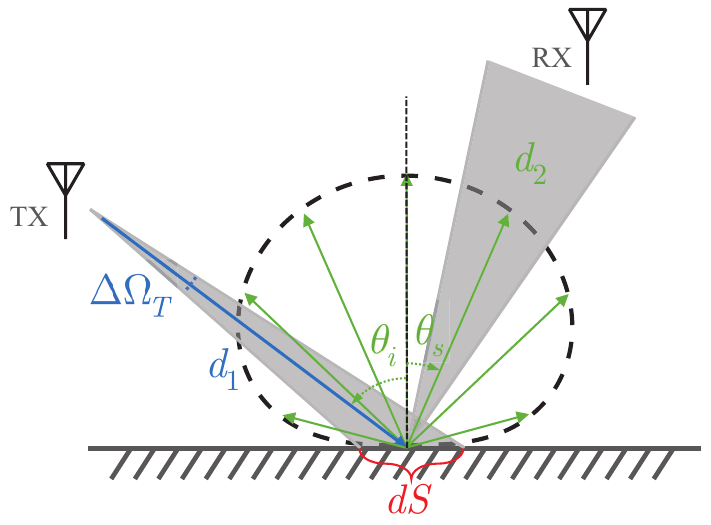}
        \caption{The scattering model.}
        \label{fig::scattering}
    \end{subfigure}
    \caption{The reflection and scattering model of mmWaves.}
    \label{fig::model}
    \vspace{-0.2 cm}
\end{figure}

The reflection path can be constructed using mirror sources, as shown in Fig. \ref{fig::model}(\subref{fig::reflection}), and the received signal power is
\begin{equation} \label{eqn::reflection}
    P_R^{(r)} = \alpha _r^2 \cdot {P_T} \cdot \frac{{{\lambda ^2}}}{{(4\pi)^2 {{({d_1} + {d_2})}^2}}} ,
\end{equation}
where $P_T$ is the total transmit power and $\alpha_r$ is the reflection attenuation coefficient.
The scattering point can be viewed as a new radiation source excited by the incident ray, as shown in Fig. \ref{fig::model}(\subref{fig::scattering}). 
According to the Lambertian scattering model for the mmWave signal in \cite{measurement_degli2007}, the received signal power is
\begin{equation}
    P_R^{(s)} = \alpha _s^2 \cdot \frac{{{P_T} \cdot {\lambda ^2}}}{{16{\pi ^3}d_1^2d_2^2}} \cdot {\rm{cos}}{\theta _i} \cdot \cos {\theta _s}{\mkern 1mu} dS ,
\end{equation}
where $\theta_i$, $\theta_s$, $\alpha_s$, $dS$ are the incidence angle, the scattering angle, the scattering attenuation coefficient, and the microfacet area, respectively.
According to the conservation of energy, $\alpha_r^2+\alpha_s^2=1$ should be satisfied.

According to the SBR method, the delay $\tau_p$, the AoA $\phi_p$, and the AoD $\varphi_p$ of each path can be calculated by numerical simulation. 
Assuming that there are $N_t$ transmit antennas, $N_r$ receive antennas, and $N_c$ subcarriers, 
then the multipath channel response $\mathbf{H}$ is a tensor with dimension $N_r \times N_t \times N_c$, and its elements are given by 
\begin{equation} \label{eqn::H}
\begin{aligned}
    \mathbf{H}_{n_r,n_t,n_c} = \sum_{p}{\beta_p} &\cdot {e^{ - j2\pi ({f_0} - \frac{{{N_c}}}{2}\Delta f)\tau_p }} \cdot {e^{ - j2\pi {n_c}\Delta f\tau_p }} \\[-1em]
    &\cdot {e^{ - j2\pi {n_t}\frac{d}{\lambda }\sin \varphi_p }} \cdot {e^{j2\pi {n_r}\frac{d}{\lambda }\sin \phi_p }} ,
\end{aligned}
\end{equation}
where $f_0$, $\Delta f$, $d$ denote the central frequency, subcarrier spacing, and array element spacing, respectively.
The path gain $\beta = \sqrt{P_R/P_T}$. 
As a result, the relationship between the received symbol $r(n_r,n_c)$ at the $n_r$-th receive antenna on the $n_c$-th subcarrier 
and the transmitted symbol $s(n_t,n_c)$ at the $n_t$-th transmit antenna on the $n_c$-th subcarrier is
\begin{equation} \label{eqn::s_r}
    r(n_r,n_c)=\sum_{n_t=0}^{N_t-1}{s(n_t,n_c)} \cdot \mathbf{H}_{n_r, n_t, n_c} + z ,
\end{equation}
where $z$ is additive white Gaussian noise. 
The waveforms from different transmitter antennas need to be orthogonal so that the receiver can separate them and estimate the channel responses, 
which can be achieved through time division.

\section{Point Target Reconstruction}\label{sec::params}
\subsection{Scattering Points Detection}
The periodogram is a commonly used spectral analysis technique in signal processing. 
According to (\ref{eqn::H}), we can obtain the spectrum of $\mathbf{H}$, which is
\begin{equation} \label{eqn::F}
\begin{aligned}
    \mathbf{F}_{i_\phi, i_\varphi,i_\tau} = \sum_{n_c=0}^{N_c-1} \sum_{n_r=0}^{N_r-1} \sum_{n_t=0}^{N_t-1} &\mathbf{H}_{n_r, n_t, n_c} \cdot e^{-j 2\pi n_r \frac{(i_\phi-N_r/2)}{N_r}}  \\[-0.5em]
                                                                                                           &\cdot e^{j 2\pi n_t \frac{(i_\varphi-N_t/2)}{N_t}} \cdot e^{j 2\pi n_c \frac{i_\tau}{N_c}}, 
\end{aligned}
\end{equation}
and the periodogram is $\mathbf{S}_{i_\phi, i_\varphi,i_\tau} = \|\mathbf{F}_{i_\phi, i_\varphi,i_\tau}\|_2^2$. 
Equation (\ref{eqn::F}) can be efficiently implemented using fast Fourier transform (FFT) and inverse fast Fourier transform (IFFT).
Here we use $\mathcal{F}_c^{-1}(\cdot)$, $\mathcal{F}_t^{-1}(\cdot)$, $\mathcal{F}_r(\cdot)$ 
to denote the transformation in the dimensions of subcarrier frequency, transmitting antenna, and receiving antenna, respectively. 
Then (\ref{eqn::F}) can be rewriten as
\begin{equation}
    \mathbf{F} = \mathcal{F}_c^{-1} \circ \mathcal{F}_t^{-1} \circ \mathcal{F}_r (\mathbf{H}), 
\end{equation}
which transforms the channel response to the AoA, AoD and delay domains.
The estimated parameters in each periodogram index can be given by 
\begin{align}
    \phi(i_\phi) &= \arcsin \left(\frac{\lambda}{d}\left(\frac{i_\phi}{N_r} - \frac{1}{2}\right) \right) ,  \label{eqn::i_phi} \\
    \varphi(i_\varphi) &= \arcsin \left(\frac{\lambda}{d}\left(\frac{i_\varphi}{N_t} - \frac{1}{2}\right)\right) , \label{eqn::i_varphi}\\
    \tau(i_\tau) &= \frac{i_\tau}{B} , \label{eqn::i_tau}
\end{align}
where $\phi(i_\phi)$, $\varphi(i_\varphi)$, $\tau(i_\tau)$ are estimated AoA, AoD and time delay respectively, and $B$ is the system bandwidth.

Since the high computational complexity of detection on the three-dimensional tensor $\mathbf{S}$, 
we simplify this process by utilizing the property of signal propagation. 
Since only the first-order reflection and scattering are considered, in the case of dual-BS, 
the target point can be located based on AoA and AoD, and they have a one-to-one correspondence.
Therefore, we can first calculate the periodogram in the AoA domain, 
\begin{equation} \label{eqn::S_AoA}
    \mathbf{S}^\phi_{i_\phi} = \frac{1}{{{N_t}{N_c}}}\sum\limits_{{n_t} = 0}^{{N_t} - 1} {\sum\limits_{{n_c} = 0}^{{N_c} - 1} {\left\| \left.\mathcal{F}_r(\mathbf{H}_{n_r,n_t,n_c})\right|_{i_\phi} \right\|_2^2} } .
\end{equation}
A one-dimensional constant false alarm rate (CFAR) detection can be applied in AoA domain, 
and we can obtain a detection map $D^{\phi}(i_\phi) \in \{0,1\}$.
Then we calculate the periodogram in the AoA-AoD domain, 
\begin{equation} \label{eqn::S_AoA_AoD}
    \mathbf{S}^{\phi ,\varphi}_{i_\phi,i_\varphi} = \frac{1}{{{N_c}}}\sum\limits_{{n_c} = 0}^{{N_c} - 1} {\left\| \left.{{\cal F}_t^{ - 1} \circ {{\cal F}_r}(\mathbf{H}_{n_r,n_t,n_c})} \right|_{i_\phi, i_\varphi} \right\|_2^2} .
\end{equation}
Besides, the AoD detection can be accomplished by searching the spectral peak on each row where the AoA is detected,
\begin{equation} \label{eqn::detect_AoD}
    \widehat{i_\varphi}(\widehat{i_\phi}) = \arg \max \limits_{{i_\varphi }} \mathbf{S}^{\phi ,\varphi}_{\widehat{i_\phi},i_\varphi},\quad \left\{ \left. \widehat{i_\phi} \right| {D^\phi}(\widehat{i_\phi}) = 1 \right\}.
\end{equation}
In the case of single-BS, there can be at most one path at a specific AoA. 
Therefore, the detection of AoA is performed first, and then detect the time delay by searching for the spectral peak on the corresponding rows in the AoD-delay domain. 
The periodogram in the AoA-delay domain is 
\begin{equation} \label{eqn::S_AoA_delay}
    \mathbf{S}^{\phi ,\tau}_{i_\phi,i_\tau} = \frac{1}{{{N_t}}}\sum\limits_{{n_t} = 0}^{{N_t} - 1} {\left\| \left.{{\cal F}_c^{ - 1} \circ {{\cal F}_r}(\mathbf{H}_{n_r,n_t,n_c})} \right|_{i_\phi, i_\tau} \right\|_2^2} ,
\end{equation}
and the time delay detection is
\begin{equation} \label{eqn::detect_AoD}
    \widehat{i_\tau}(\widehat{i_\phi}) = \arg \max \limits_{{i_\tau }} \mathbf{S}^{\phi ,\tau}_{\widehat{i_\phi},i_\tau},\quad \left\{ \left. \widehat{i_\phi} \right| {D^\phi}(\widehat{i_\phi}) = 1 \right\} .
\end{equation}

According to (\ref{eqn::i_phi})-(\ref{eqn::i_tau}), the periodogram-based detection has a limited resolution, which greatly affects the imaging accuracy.
Subspace methods such as multiple signal classification (MUSIC) and estimation of signal parameters via rational invariance techniques (ESPRIT) can be combined to achieve super-resolution estimation of target parameters. 

The MUSIC algorithm first performs eigenvalue decomposition on the covariance matrix $\mathbf{R_x}$ of the $M$-dimensional observation signal $\mathbf{x}$. 
Assuming that the number of targets is $K$, then the eigenvectors corresponding to the first $K$ eigenvalues form the signal subspace $\mathbf{U}_s = [\mathbf{u}_1,\mathbf{u}_2,\cdots,\mathbf{u}_K]$, 
and the remained $M-K$ eigenvectors form the noise subspace $\mathbf{U}_n = [\mathbf{u}_{K+1},\mathbf{u}_{K+2},\cdots,\mathbf{u}_{M}]$. 
Based on the orthogonality between the target's frequency vector $\mathbf{a}(\omega_k) = [1,e^{j\omega_k},\cdots,e^{j(M-1)\omega_k}]^T$ and the noise subspace $\mathbf{U}_n$, 
the estimation of $\omega_1,\omega_2,\cdots,\omega_K$ can be achieved by searching for $K$ peaks of the following pseudo-spectrum function, 
\begin{equation} \label{eqn::MUSIC_spectrum}
    P_{\rm MUSIC}(\omega) = \frac{1}{\mathbf{a}^H(\omega) \mathbf{U}_n \mathbf{U}^H_n \mathbf{a}(\omega)}.
\end{equation}

Here, we use the MUSIC algorithm to estimate the AoD and time delay corresponding to each detected AoA, replacing the original process of searching for spectral peaks.
Specifically, we first compute the covariance matrices $\mathbf{R}_t^{\phi}$ and $\mathbf{R}_c^{\phi}$, 
\begin{equation} \label{eqn::R_phi}
    \mathbf{R}_t^{\phi} = \frac{1}{N_t} \mathbf{H}_{\phi} \mathbf{H}_{\phi}^H, \enspace
    \mathbf{R}_c^{\phi} = \frac{1}{N_c} \mathbf{H}_{\phi}^T \mathbf{H}_{\phi}^*, 
\end{equation}
where $\mathbf{H}_{\phi} = \mathcal{F}_r(\mathbf{H}) |_{\widehat{i_\phi}} \in \mathbb{C}^{N_t \times N_c}$ is the two-dimension slice of the channel response at the detected AoA index $\widehat{i_\phi}$. 
Then, the MUSIC algorithm with target number $K=1$ can be applied to estimate the AoD and time delay for each $\widehat{i_\phi}$.

\subsection{Reflection Points Detection} \label{subsec::Reflection_Points_Detection}
Unlike the clustering characteristics of scattering paths, reflection paths have sparsity in space and usually have stronger signal power. 
It can be proven that there is at most one reflection path formed between a transceiver pair in the scenario of a single convex polygon. 
Therefore, we can determine the existence of the reflection path based on the strongest signal power in the periodogram $\mathbf{S}$.  
We use $\mathbf{R}_r$, $\mathbf{R}_t$, $\mathbf{R}_c$ to denote the covariance matrices in the domain of receive antenna, transmit antenna and subcarrier, respectively, where
\begin{equation}
    \mathbf{R}_r = \frac{1}{N_t N_c}\sum_{n_t=0}^{N_t-1} \sum_{n_c=0}^{N_c-1} {\mathbf{H}_{:,n_t,n_c} \mathbf{H}_{:,n_t,n_c}^H}, 
\end{equation}
and $\mathbf{R}_t$, $\mathbf{R}_c$ can be calculated in the same way. 
If a reflection path exists, we can use the MUSIC algorithm to estimate its AoA, AoD, and time delay separately using these three covariance matrices.
Based on the path parameters, it is easy to obtain the position of the reflection point and the direction of the reflection surface.

\section{Target Shape Reconstruction}\label{sec::shape}
Under the assumption of a single convex polygon target, the task of shape reconstruction is to fit a polygon based on detected points. 
However, the scattering and reflection points have different features. 
It's not easy to fuse them on a point level directly. Therefore, we designed a two-step approach for shape reconstruction. 
In the first step, we propose the HT-PCA-TSR algorithm for preliminary target shape fitting based on scattering points. 
In the second step, the shape is refined by combining reflection paths.

\subsection{The HT-PCA-TSR Algorithm}
The Hough transform is a classic algorithm for line detection in image processing. 
For any line $l$ passing through a fixed point $(x_0,y_0)$, its parameter $\{\rho, \theta\}$ satisfies
\begin{equation}
    l: \rho = x_0 \cos{\theta} + y_0 \sin{\theta} .
\end{equation}
Therefore, all lines passing through a certain fixed point correspond to a sine curve in the $\rho-\theta$ space. 
By discretizing the $\rho-\theta$ space into $N_{\rho-\theta}$ grids of size $\Delta \rho \times \Delta \theta$ and counting the number of sine curves passing through each grid, 
we can obtain the number of points on each line. 

However, the line detection only based on the Hough transform of scattering points is not well. 
On the one hand, the discretized parameter space limits the detection accuracy. 
On the other hand, the grid parameters do not correspond to the best-fit line for the points contained within the grid.
Obviously, we can use the principal component analysis (PCA) to calculate the best-fit line $l(\mathcal{P})$ of the point set $\mathcal{P}$.

\begin{figure}[!htbp]
    \vspace{-0.3 cm}
    \centering
    \begin{subfigure}[b]{0.47\linewidth}
        \centering
        \includegraphics[width=\linewidth]{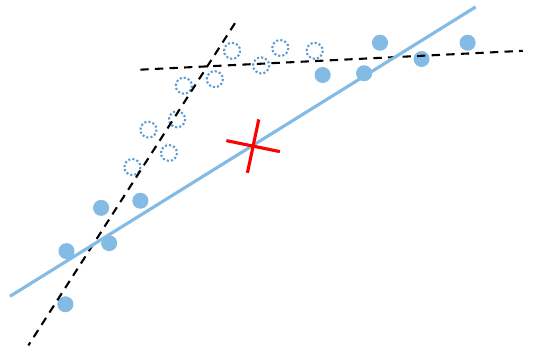}
        \caption{The distribution of points along the line is non-uniform.}
        \label{fig::invalid_line_1}
    \end{subfigure}
    \hfill
    \begin{subfigure}[b]{0.47\linewidth}
        \centering
        \includegraphics[width=\linewidth]{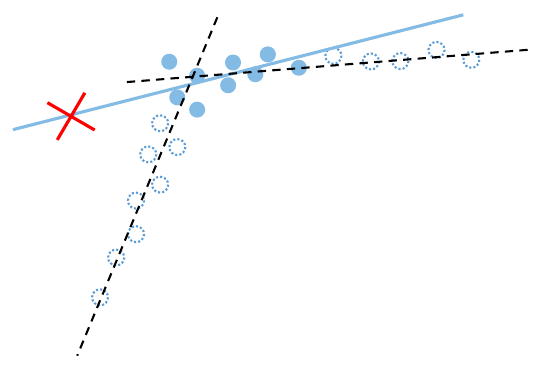}
        \caption{The variance of points in the line direction is relatively small.}
        \label{fig::invalid_line_2}
    \end{subfigure}
    \caption{Two cases of invalid line fitting.}
    \label{fig::invalid_line}
    \vspace{-0.4 cm}
\end{figure}

We need to make a further judgment on the validity of the fitting results. 
A proper fitting line should satisfy the following two conditions: 
(\romannumeral1) points should be approximately uniformly distributed in the line direction; 
(\romannumeral2) the variance of points along the line should be significantly greater than that in the perpendicular direction.
The two cases of invalid line fitting are shown in Fig. \ref{fig::invalid_line}. 
For the assessment of projection uniformity, we consider the following hypothesis test problem: 
\begin{equation} \label{eqn::uniform_dist}
    H_0: \operatorname{proj}(\mathcal{P}, l(\mathcal{P})) \sim \mathcal{U}(a,b)
\end{equation}
where we use $\operatorname{proj}(\mathcal{P}, l)$ to denote the projection of points $\mathcal{P}$ on line $l$.
For the assessment of variance, we use 
\begin{equation} \label{eqn::variance_ratio}
    \sigma_1(\mathcal{P}) / \sigma_2(\mathcal{P}) \geq \gamma_v
\end{equation}
to compare the variance of $\mathcal{P}$ in two directions, where $\sigma_1(\mathcal{P})$ and $\sigma_2(\mathcal{P})$ are the variances of the major and minor components obtained by PCA algorithm, and $\gamma_v$ is a threshold.

In addition, not every point set obtained by the Hough transform corresponds to an independent line. 
There may be overlap between many point sets, so merging some point sets is needed. 
We first sort the grids obtained by the Hough transform in descending order based on the number of points in each grid, 
using $\mathcal{P}_{i}^{\rm{h}}$ to denote the points in the $i$-th grid. 
Then, we sequentially examine the relationship between $\mathcal{P}_{i}^{\rm{h}}$ and the point set corresponding to each fitted line, 
using $\mathcal{P}_{k}^{\rm{s}}$ to denote the points belonging to the current $k$-th fitted line.
If $\mathcal{P}_{i}^{\rm{h}}$ and $\mathcal{P}_{k}^{\rm{s}}$ satisfy the condition of 
\begin{equation} \label{eqn::union_line_condition}
    \frac{| \mathcal{P}_{i}^{\rm{h}} \cap \mathcal{P}_{k}^{\rm{s}} |}{| \mathcal{P}_{k}^{\rm{s}} |} \geq \gamma_l \enspace 
    \text{and} \enspace 
    | \operatorname{ang}(l(\mathcal{P}_{i}^{\rm{h}})) - \operatorname{ang}(l(\mathcal{P}_{k}^{\rm{s}})) | \leq \epsilon_\theta ,
\end{equation}
we determine that the corresponding lines of these two sets of points have sufficient similarity and merge $\mathcal{P}_{i}^{\rm{h}}$ into $\mathcal{P}_{k}^{\rm{s}}$, 
where $\operatorname{ang}(\cdot)$ calculates the line's direction. 
If the overlap between $\mathcal{P}_{i}^{\rm{h}}$ and all $\mathcal{P}_{k}^{\rm{s}}$ is sufficiently low, 
that is
\vspace{-0.05cm}
\begin{equation} \label{eqn::new_line_condition}
    \frac{| \mathcal{P}_{i}^{\rm{h}} \cap \mathcal{P}_{k}^{\rm{s}} |}{| \mathcal{P}_{k}^{\rm{s}} |} \leq \gamma_u, \enspace \forall k, 
\vspace{-0.05cm}
\end{equation}
then we consider $\mathcal{P}_{i}^{\rm{h}}$ as a new fitted line. 
The $\gamma_l$, $\gamma_u$, $\epsilon_\theta$ are all pre-set parameters, where $\gamma_l < \gamma_u$.
We summarize the proposed HT-PCA-TSR algorithm in Algorithm \ref{algo::HT-PCA-TSR}.

{\small
\begin{algorithm}[htbp]
\caption{The HT-PCA-TSR Algorithm}\label{algo::HT-PCA-TSR}
\begin{algorithmic}[1]
\Require Given the scattering points $\mathcal{P}_{\rm{s0}}$. 
\Ensure The lines of output target $\mathcal{L}_{\rm{s}} = \{l^{\rm{s}}_k|k=1,\cdots,K\}$ and the points for all fitted line $\mathcal{P}_{\rm{s}}=\{\mathcal{P}^{\rm{s}}_k|k=1,\cdots,K\}$. 
\State \textbf{Initialization}: Set the edge number of the output target $K = 0$. Set $\mathcal{L}_{\rm{s}} = \varnothing$, $\mathcal{P}_{\rm{s}} = \varnothing$. 
\State Perform Hough transform on $\mathcal{P}_{\rm{s0}}$ and sort the $N_{\rho-\theta}$ subsets in descending order according to their size $|\mathcal{P}_{i}^{\rm{h}}|$. 
\For{$i = 1,\cdots,N_{\rho-\theta}$}
    \State $l_i = l(\mathcal{P}_{i}^{\rm{h}})$. 
    \State Put $\mathcal{P}_{i}^{\rm{h}}$ into (\ref{eqn::uniform_dist}) and (\ref{eqn::variance_ratio}) to validate $l_i$. 
    \If{$l_i$ is valid}
        \State $\text{flag} = \text{true}$.
        \For{$k = 1,\cdots,K$}
            \If{(\ref{eqn::new_line_condition}) is false}
                \State $\text{flag} = \text{false}$.
            \EndIf
            \If{(\ref{eqn::union_line_condition}) is true}
                \State $\mathcal{P}_{k}^{\rm{s}} = \mathcal{P}_{k}^{\rm{s}} \cup \mathcal{P}_i^{\rm{h}} \text{,} \enspace l_k^{\rm{s}} = l(\mathcal{P}_k^{\rm{s}})$.
            \EndIf
        \EndFor
        \If{flag is true}
            \State $\mathcal{P}_{\rm{s}}=\mathcal{P}_{\rm{s}} \cup \{\mathcal{P}_{i}^{\rm{h}}\} \text{,}\enspace \mathcal{L}_{\rm{s}} = \mathcal{L}_{\rm{s}} \cup \{l_i\} \text{,}\enspace K = K+1$.
        \EndIf
    \EndIf
    
\EndFor
\end{algorithmic}
\end{algorithm}
}

\vspace{-0.2cm}

\subsection{Target Shape Refinement}
According to Section \ref{subsec::Reflection_Points_Detection}, we have been able to obtain the AoA, AoD and time delay of the reflection path, 
which can be used to calculate the line $l^{\rm{r}}$ of the reflection surface.
In a high signal-to-noise ratio (SNR) situation, the detected reflection surface is often the repetitive detection of a certain scattering surface. 
At this time, $l^{\rm{r}}$ is similar to a particular scattering surface.
First, we find the scattering surface $l^{\rm{s}}_{\tilde{k}}$ in $\mathcal{L}_{\rm{s}}$ that is closest to $l^{\rm{r}}$, 
\begin{equation}
    \tilde{k} = \arg \min_{k} {d\left( \mathcal{P}^{\rm{s}}_{k}, l^{\rm{r}} \right)},
\end{equation}
where $d(\mathcal{P},l)$ denotes the average distance from the points in the set $\mathcal{P}$ to the line $l$. 
Then, we compare $\sigma_{\rm{s}} = d(\mathcal{P}^{\rm{s}}_{\tilde{k}}, l^{\rm{s}}_{\tilde{k}})$ and $\sigma_{\rm{r}} = d(\mathcal{P}^{\rm{s}}_{\tilde{k}}, l^{\rm{r}})$. 
If $\sigma_{\rm{r}} \leq \gamma_s \cdot \sigma_{\rm{s}}$, we consider that $l^{\rm{r}}$ is a repetitive detection of $l^{\rm{s}}_{\tilde{k}}$ 
and perform a parameter fusion of these two lines to get a new line $l^{\rm{sr}}: a_r x + b_r y + c_{sr} = 0$.
The $a_r$, $b_r$ are the corresponding parameters of line $l^{\rm{r}}$, and $c_{sr}$ is
\begin{equation} \label{eqn::opt_c}
    c_{sr} = \arg \min_{c} \{ \lambda_{r}\cdot d^2(p^{\rm{r}},l^{\rm{sr}}) + (1-\lambda_{r}) \cdot d^2(\mathcal{P}^{\rm{s}}_{\tilde{k}}, l^{\rm{sr}}) \} ,
\end{equation}
where $p^{\rm{r}}=(x_r,y_r)$ is the reflection point and $\lambda_r$ is a weighting coefficient. 
Thanks to the surface orientation contained in the reflection path parameters, the direction of the fusion result $l^{\rm{sr}}$ is more accurate than $l^{\rm{sr}}_{\tilde{k}}$.

In a low SNR situation, the noise introduced by the reflection path can mask the surrounding scattering signals, leading to the miss detection of the surface.
In this case, $l^{\rm{r}}$ will not similar to any scattering surface, which can be identified by $\sigma_{\rm{r}} > \gamma_s \cdot \sigma_{\rm{s}}$. 
To compensate for the missing surface, we incorporate $l^{\rm{r}}$ as a new edge into $\mathcal{L}_{\rm{s}}$, ensuring the integrity of the reconstructed shape.

\section{Simulation Results}\label{sec::numerical_results}
We simulate the scene containing a single convex polygon object, and the system parameters are shown in Table \ref{tab::sim_params}. 
We use the following threshold parameters: $\gamma_v = 5$, $\gamma_l = 0.2$, $\gamma_u = 0.5$, $\epsilon_\theta = 1^\circ $, $\gamma_s = 3$.

\begin{table}[h]
\centering
\caption{Simulation parameter settings.}
\begin{tabular}{|l|l|}
\hline
\textbf{Parameter}                 & \textbf{Value}          \\ \hline
Center frequency ($f_0$)           & 28GHz                   \\ \hline
System bandwidth ($B$)             & 1GHz                    \\ \hline
Subcarrier number ($N_c$)          & 256                     \\ \hline
Antenna number of BS ($N_t,N_r$)   & 128,128                 \\ \hline
\end{tabular}
\label{tab::sim_params}
\end{table}

\begin{figure}[!t]
    \centering
    \begin{subfigure}[b]{0.49\linewidth}
        \centering
        \includegraphics[width=\linewidth]{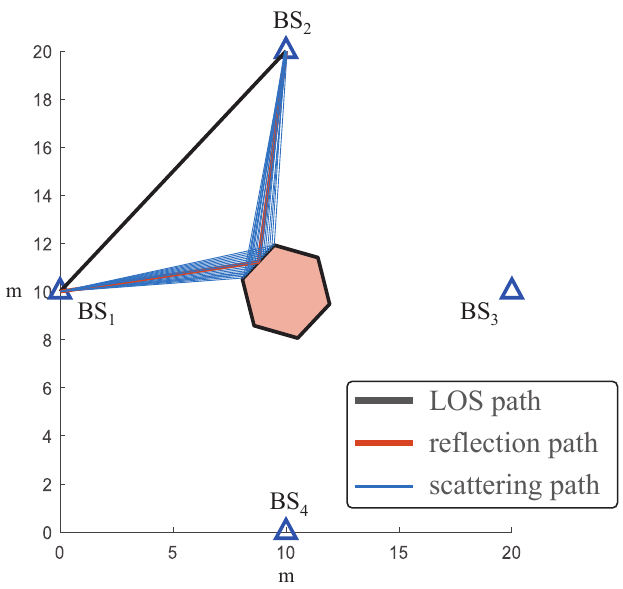}
        \caption{The channel simulation result.}
        \label{fig::RT_1}
    \end{subfigure}
    \begin{subfigure}[b]{0.49\linewidth}
        \centering
        \includegraphics[width=\linewidth]{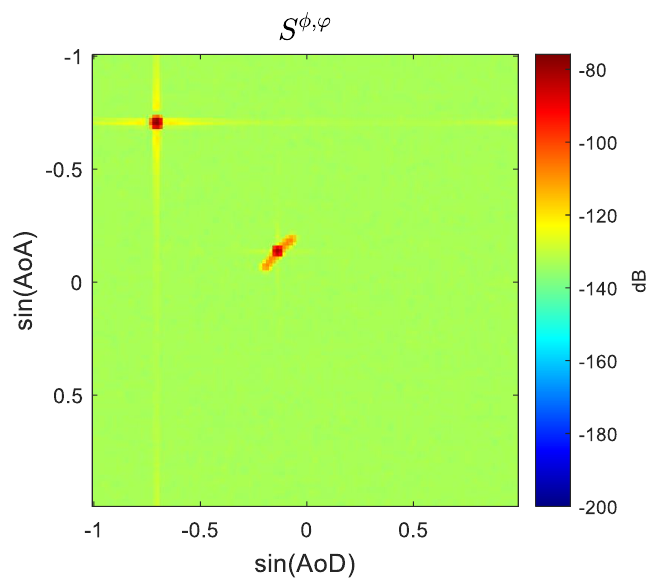}
        \caption{The periodogram $\mathbf{S}^{\phi ,\varphi}$ of $\mathbf{H}$.}
        \label{fig::RT_3}
    \end{subfigure}
    \caption{The channel simulation and periodogram analysis.}
    \label{fig::RT}
    \vspace{-0.2 cm}
\end{figure}

Fig. \ref{fig::RT}(\subref{fig::RT_1}) shows the channel simulation result based on ray-tracing in the dual-BS case, which includes scattering and reflection paths. 
The simulation results for the single-BS case are similar to this.
In practical applications, it is necessary to use windowing to suppress the interference of strong sidelobes on target detection, 
so we use $\tilde{\mathbf{H}} = \mathbf{H} \odot \mathbf{W}$ instead of $\mathbf{H}$ for the periodogram processing, 
where $\mathbf{W}$ is a Hamming window function and $\odot$ is the Hadamard product. 
Fig. \ref{fig::RT}(\subref{fig::RT_3}) shows the periodogram in the AoA-AoD domain of the simulated channel at $\text{SNR}=\text{20dB}$. 

\begin{figure} [!t] 
\centering
\includegraphics[width=\linewidth]{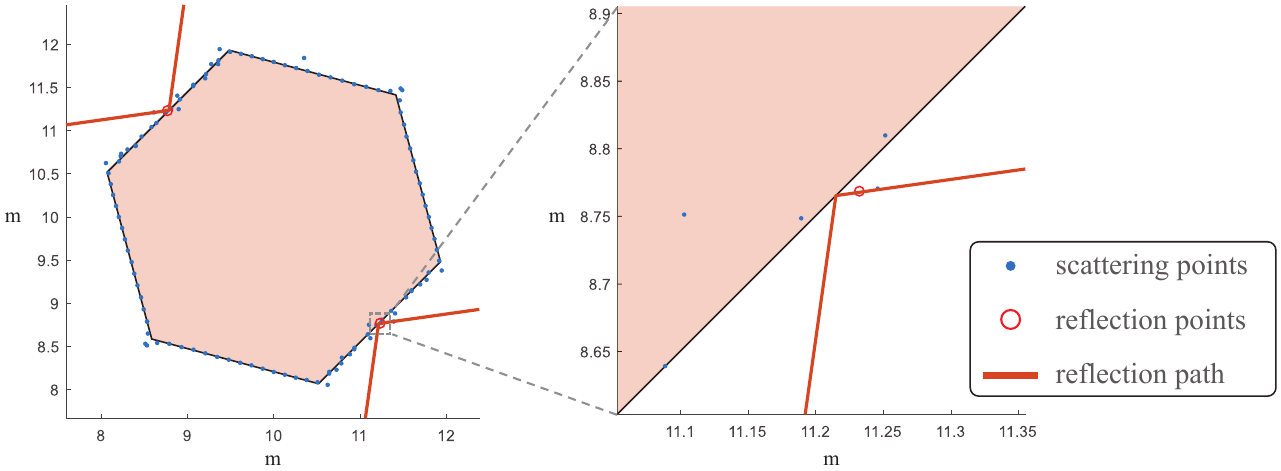}
\vspace{-0.4 cm}
\caption{Scattering and reflection points detection.}
\label{fig::points_detection}
\vspace{-0.4 cm}
\end{figure}

Fig. \ref{fig::points_detection} shows the scattering and reflection points detected by the combination of the periodogram and MUSIC algorithm. 
It can be seen that the detected points are mainly distributed on the object's surface. 
However, due to the sidelobes of the reflection signal, the detection error of the scattering points around it will be relatively large.

\begin{figure*}[!t]
    \centering
    \begin{subfigure}[t]{0.28\linewidth}
        \centering
        \includegraphics[width=\linewidth]{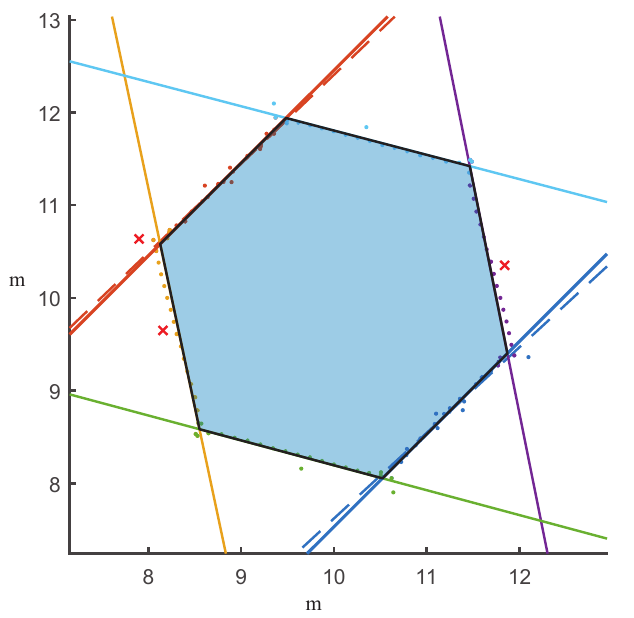}
        \caption{Fusion of reflection and scattering surface at SNR=20dB. 
        The dashed and solid lines represent the surface before and after the refinement, respectively.}
        \label{fig::shape_1}
    \end{subfigure}
    \hfill
    \begin{subfigure}[t]{0.28\linewidth}
        \centering
        \includegraphics[width=\linewidth]{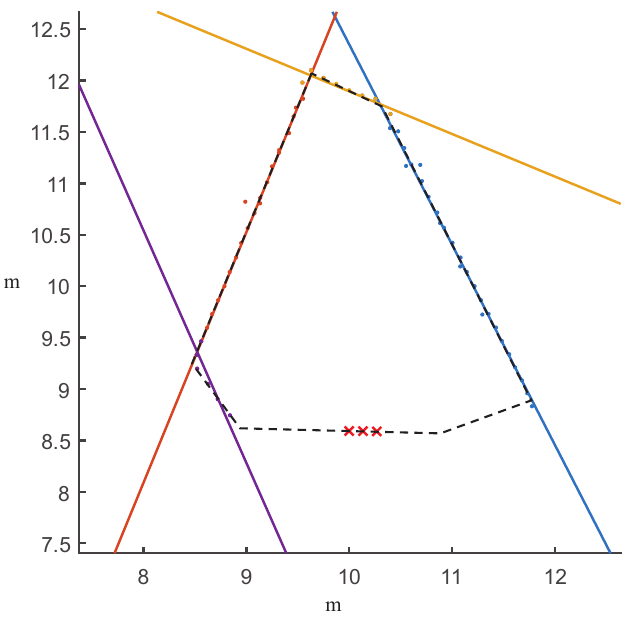}
        \caption{Miss detection of the surface at SNR=10dB. 
        The black dashed line represents the ground truth shape.}
        \label{fig::shape_2}
    \end{subfigure}
    \hfill
    \begin{subfigure}[t]{0.28\linewidth}
        \centering
        \includegraphics[width=\linewidth]{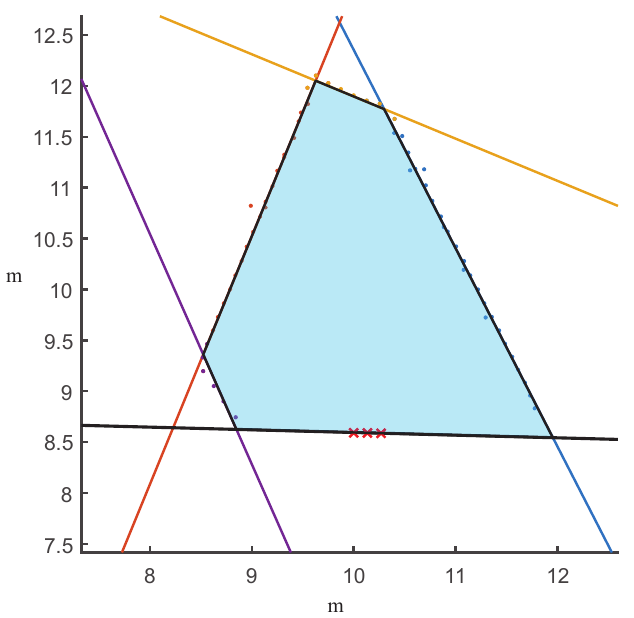}
        \caption{Shape completion by combining reflection information. The bold black solid line represents the completed surface.}
        \label{fig::shape_3}
    \end{subfigure}
    \vspace{-0.1 cm}
    \caption{Target shape reconstruction results.}
    \label{fig::shape}
    \vspace{-0.4 cm}
\end{figure*}

Fig. \ref{fig::shape} shows the target shape refinement in the presence of the reflection path. 
Fig. \ref{fig::shape}(\subref{fig::shape_1}) illustrates adjusting the fitted surface when the SNR is relatively high. 
When the SNR is low, the masking effect of a strong reflection path can cause missed detection of scattering points in the surrounding area, as shown in Fig. \ref{fig::shape}(\subref{fig::shape_2}). 
Using the reflection point information, we can complete the missing surface, as shown in Fig. \ref{fig::shape}(\subref{fig::shape_3}).

To estimate the performance of scattering points detection, we calculate the MSE, 
\vspace{-0.2cm}
\begin{equation}
\vspace{-0.05cm}
    \operatorname{MSE}({{\cal P}_{\rm{s0}}}, s_{\rm{tgt}}) = \frac{1}{{\left| {{{\cal P}_{\rm{s0}}}} \right|}}\sum\limits_{i = 1}^{\left| {{{\cal P}_{\rm{s0}}}} \right|} {{{\left( {\mathop {\min }\limits_{l \in {\mathop{\rm E}\nolimits} (s_{\rm{tgt}})} d({{\cal P}_{\rm{s0},i}},l)} \right)}^2}} ,
\end{equation}
where $\operatorname{E}(s_{\rm{tgt}})$ is the edge set of the ground truth shape $s_{\rm{tgt}}$.
By comparing periodogram-based scattering points detection (Per-SPD) and periodogram-MUSIC-based scattering points detection (Per-MUSIC-SPD), 
we can find that combining subspace method effectively improves the sensing accuracy of scattering points, which is shown in Fig. \ref{fig::shape}(\subref{fig::shape_1}). 
As the SNR increases, the MSE of the reconstructed results obtained by the two algorithms shows a downward trend. 
However, for Per-SPD, the sidelobes of the reflection signal can cause more false detections at high SNRs, leading to additional errors.

We test the shape reconstruction performance from the surface direction error and the shape close rate. 
We perform shape reconstruction using the HT-PCA-TSR algorithm on the outputs of Per-PSD and Per-MUSIC-PSD, and the shape refinement is based on the outputs of Per-MUSIC-PSD. 
The test results are shown in Fig. \ref{fig::shape}(\subref{fig::shape_2}) and \ref{fig::shape}(\subref{fig::shape_3}). 
Combining the reflection point information can significantly improve the accuracy of surface direction and ensure the integrity of reconstructed shapes under low SNR conditions.

\begin{figure*}[!t]
    \centering
    \begin{subfigure}[t]{0.32\linewidth}
        \centering
        \includegraphics[width=\linewidth]{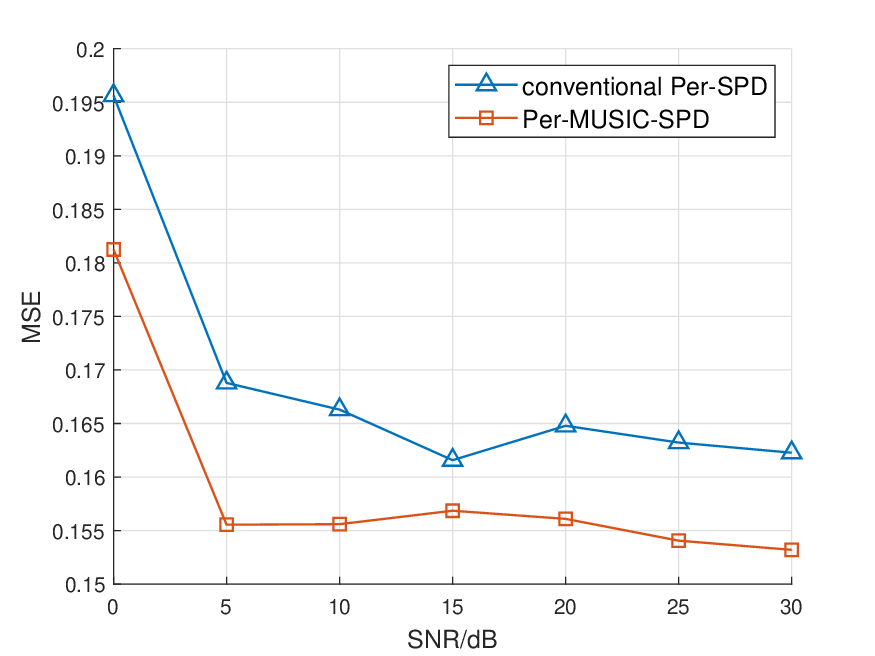}
        \caption{The MSE of scattering points detection.}
        \label{fig::scatter_MSE}
    \end{subfigure}
    \hfill
    \begin{subfigure}[t]{0.32\linewidth}
        \centering
        \includegraphics[width=\linewidth]{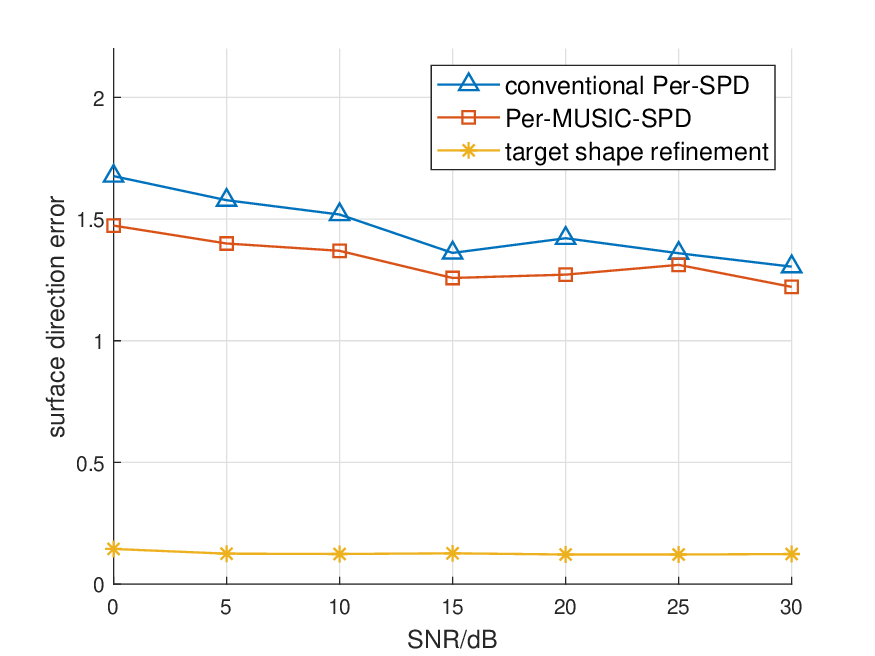}
        \caption{The sensing error of surface direction.}
        \label{fig::direction_error}
    \end{subfigure}
    \hfill
    \begin{subfigure}[t]{0.32\linewidth}
        \centering
        \includegraphics[width=\linewidth]{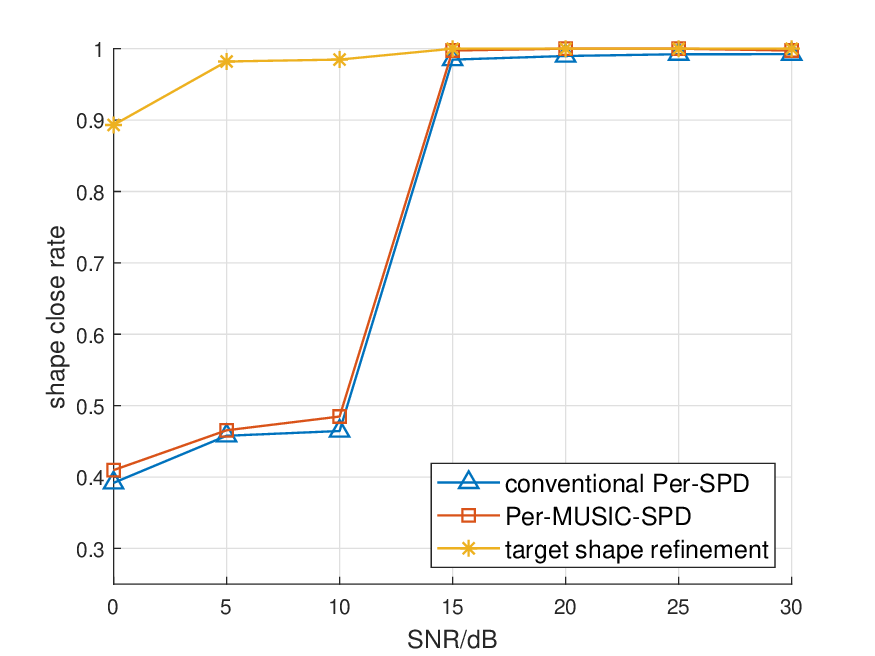}
        \caption{The close rate of reconstructed shapes.}
        \label{fig::close_rate}
    \end{subfigure}
    \caption{The performance of scattering point detection and shape reconstruction under different SNRs.}
    \label{fig::performance}
    \vspace{-0.4 cm}
\end{figure*}

\section{Conclusions}\label{sec::con}
In this paper, we propose a scheme for target shape sensing by fusing mmWave reflection and scattering points. 
First, we model the propagation characteristics of millimeter waves and construct channels using ray-tracing. 
Then, we design a point detection algorithm by combining the periodogram and the subspace method. 
Next, we propose the HT-PCA-TSR algorithm and a target shape refinement method to achieve shape reconstruction of polygon targets. 
Numerical simulation results validate the effectiveness of the proposed algorithms. 
Future directions include the sensing of nonconvex polygon target and multiple targets with the ray tracing channel model.

\section*{Acknowledgment}
This work was supported in part by National Natural Science Foundation of China under Grant U20A20158, National Key R\&D Program of China under Grant 2020YFB1807101 and 2018YFB1801104, Ministry of Industry and Information Technology under Grant TC220H07E, and Zhejiang Provincial Key R\&D Program under Grant 2023C01021.

\small
\printbibliography

\end{document}